\documentclass{article}
\usepackage{spconf,amsmath,graphicx}
\usepackage{hyperref}
\hypersetup{
  colorlinks   = true, %Colours links instead of boxes
  urlcolor     = black, %Colour for external hyperlinks
  linkcolor    = black, %Colour of internal links
  citecolor   =  black %Colour of citations
}
\usepackage{orcidlink}
\usepackage{booktabs}

\usepackage{xurl}

\usepackage{amssymb}

\usepackage{booktabs}
\usepackage{colortbl}

\usepackage{tikz}
\usetikzlibrary{backgrounds, fit}

\usepackage{mathtools}
\usetikzlibrary{calc,positioning}

\usepackage{subcaption}

\usepackage{diagbox}

\usepackage{threeparttable}
\usepackage{caption}

\title{Cross-Codec robustness in speech watermarking}
\name{
Olli-Pekka Pauna$^{1}$\sthanks{This work was a part of Finland's Ministry of Education and Culture’s Doctoral Education Pilot under Decision No. VN/3137/2024-OKM-6 (The Finnish Doctoral Program Network in Artificial Intelligence, AI-DOC).}\orcidlink{0009-0009-0286-2388}\qquad
Xin Wang$^{2}$\sthanks{Xin Wang is supported by JST PRESTO (No. JPMJPR23P9), Japan.}\orcidlink{0000-0001-8246-0606}\qquad
Lauri Juvela$^{1}$
\sthanks{This research was funded by the Research Council of Finland (grant no. 378123).}
\orcidlink{0000-0002-2201-103X}
}

\address{
$^{1}$ Department of Information and Communications Engineering,
Aalto University, Espoo, Finland \\
$^{2}$ National Institute of Informatics, Tokyo, Japan
}
\begin{document}
\ninept
\maketitle
\begin{abstract}
Generative speech models synthesize audio that is difficult to distinguish from human recordings. Synthetic speech can be watermarked as a proactive disclosure method, but watermarking models remain vulnerable to resynthesis attacks using neural audio codecs. This article investigates speech watermarking robustness across neural codec and digital signal processing attacks, when specific neural codecs are used for channel augmentation during training. We also present a novel zero-bit watermarking system that uses Descript Audio Codec as the generator. Our results demonstrate that the choice of augmentation codec significantly influences robustness generalization. Experiments show that augmentation with low-bitrate, high-quality neural codecs---Mimi and SNAC---results in promising robustness against unseen neural codecs and traditional attacks with limited quality impact.
\end{abstract}
\section{Introduction}

Text-to-speech models now synthesize recordings that highly resemble human recordings in a zero-shot way~\cite{naturalspeech3, valle, chen-etal-2025-f5}. Open-sourcing of pretrained models makes voice cloning accessible to researchers and software engineers~\cite{ chen-etal-2025-f5, casanova2022yourtts}. To counteract potential misuse, active disclosure methods such as watermarking are needed and increasingly mandated by legislation, most prominently the European Union AI Act~\cite{european_commission_technical_2026}. Digital watermarking attempts to embed perceptually hidden but algorithmically readable information within digital content~\cite{watermarking_is_not}. Encoded payloads can range from longer bit strings, which can be used to identify, e.g., the specific generative model~\cite{yu2022responsible}, to zero-bit watermarks~\cite{juvela2024-collaborative-watermarking}, which are used for binary detection. 

Watermarking should be imperceptible enough not to hinder the beneficial applications of generative speech models while remaining robust against removal attacks. Deep-learning-based watermarking models can be trained to be robust against distortions with training-time channel augmentation~\cite{juvela2024-collaborative-watermarking, juvela2025audio,timbrewatermarking-ndss2024, audioseal, chen2023wavmark}. In these approaches, a generator model is trained to watermark audio so that, after passing through channel augmentations such as filtering or codec resynthesis, the watermark can still be decoded by a detector. Although recent models survive various classical DSP attacks while maintaining competitive audio quality, the same models remain vulnerable to multiple unseen neural-network-based attacks, such as neural codec resynthesis, vocoders, and denoising~\cite{ juvela2025audio, timbrewatermarking-ndss2024, audioseal,chen2023wavmark, shallow, ozer25_interspeech, audiomarkbench}.

Neural audio codecs are trained to compress, quantize, and reconstruct audio in an end-to-end fashion utilizing adversarial and reconstruction losses~\cite{soundstream, encodec, dac}. They have potential applications ranging from audio transmission in bandwidth-limited environments to generative modeling. Furthermore, they are relatively light-weight, provide high-quality reconstructions through a narrow bottleneck and are often open-sourced. This makes them an enticing attack option for watermark removal. 

Channel augmentation with neural audio codecs has shown promising generalization against traditional DSP codecs~\cite{juvela2025audio}. However, vulnerability to unseen neural codecs has been observed even when models are trained to be robust against individual neural codecs~\cite{shallow}. Neural codecs differ in architecture and training details, which leads to different reconstruction behavior. Ideally, we want to find an augmentation recipe that generalizes across multiple unseen neural codecs.

The contributions of this work are threefold. Firstly, we present a new watermarking system that uses the Descript Audio Codec (DAC)~\cite{dac} as the watermark generator with high audio quality. Secondly, we demonstrate that the choice of neural audio codec for watermarking channel augmentation significantly influences robustness generalization. Third, we show that joint augmentation with recent very efficient neural audio codecs---Mimi and SNAC---achieves promising robustness across unseen neural codecs and DSP attacks with limited quality impact. Furthermore, we open-source our channel augmentation implementations. % for the audio watermarking research community.

\vspace{-6pt}
\section{Related Work}
\vspace{-4pt}

Standard neural audio codecs, such as SoundStream, EnCodec, and DAC use residual vector quantization (RVQ) for implementing the Vector Quantized-Variational AutoEncoder architecture~\cite{soundstream, encodec, dac, van_den_oord_neural_2017}. SpeechTokenizer also uses RVQ, but it disentangles speech information by utilizing a semantic distillation task on the first codebook~\cite{zhang2023speechtokenizer}. Mimi similarly utilizes  semantic distillation for its first codebook, but it discards reconstruction losses, relying on adversarial objectives~\cite{moshi}. Multi-Scale Neural Audio Codec (SNAC) introduces an extension of RVQ, where the quantizers operate at different temporal resolutions, allowing it to capture long-range and short-term acoustic details separately and its decoder utilizes stochastic noise blocks after upsampling~\cite{siuzdak2024snac}. NanoCodec uses finite scalar quantization (FSQ), which circumvents the need to learn codebook entries~\cite{nanocodec, mentzer2024finite}. Surpassing earlier models Mimi, SNAC and NanoCodec demonstrate exceptional quality at very low bitrates~\cite{moshi, siuzdak2024snac, nanocodec}.

Recently, Latent-Mark was proposed as a means to target the shared latent subspace between different neural codecs, showcasing robustness against unseen neural codecs~\cite{chen2026latent}. We hypothesize that channel augmentation with recent highly efficient codecs implicitly targets the shared latent subspace with lower-capacity models due to the narrower bottleneck.

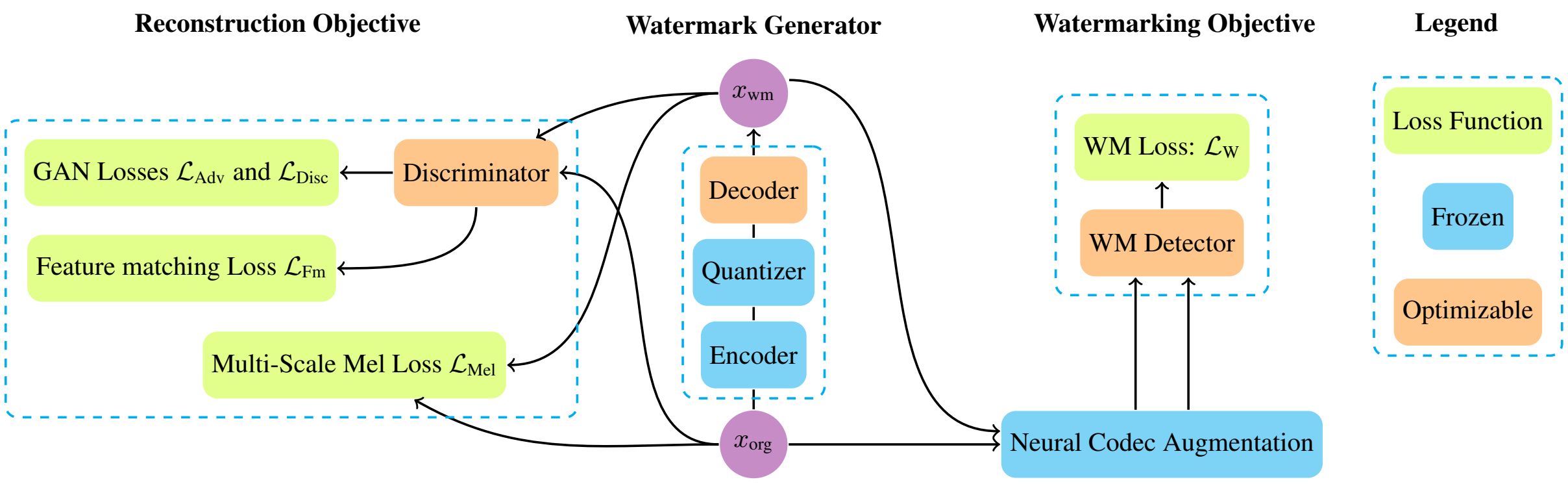
\begin{figure*}[tbp]
\centering
\resizebox{.9\textwidth}{!}{
\begin{tikzpicture}[
    ->, thick,
    data/.style={circle, fill=violet!45},
    model/.style={rectangle, fill=orange!45, minimum size=8mm, rounded corners=5pt},
    frozen/.style={rectangle, fill=cyan!45, minimum size=8mm, rounded corners=5pt},
    loss/.style={rectangle, fill=lime!45, minimum size=8mm, rounded corners=5pt},
    label/.style={font=\normalsize, font=\bfseries},
    ]
    \node[data] (xorg) {$x_\text{org}$};
    \node[frozen, above=0.75em of xorg] (encoder) {Encoder};
    \node[frozen, above=0.5em of encoder] (quantizer) {Quantizer};
    \node[model, above=0.5em of quantizer] (decoder){Decoder};
    \node[data, above=1em of decoder] (xwm) {$x_\text{wm}$};

    \node[model, left=6em of xwm, yshift=-3em] (discriminator) {Discriminator};
    \node[loss, left=8em of xorg, yshift=3em] (melloss) {Multi-Scale Mel Loss $\mathcal{L}_\text{Mel}$};

    \draw (xorg) [out=180, in=0] to (discriminator);
    \draw (xwm) [out=180, in=30] to (discriminator);

    \draw (xorg) [out=180, in=330] to (melloss);
    \draw (xwm) [out=180, in=0] to (melloss);

    \node[loss, left=2em of discriminator, yshift=0em] (gloss) {GAN Losses  $\mathcal{L}_\text{Adv}$ and $\mathcal{L}_\text{Disc}$};
    \node[loss, below=1em of gloss] (floss) {Feature matching Loss  $\mathcal{L}_\text{Fm}$};
    
    \draw (discriminator) -> (gloss);
    \draw (discriminator)[out=270, in=0] to (floss);

    \node[frozen, right=8em of xorg] (codec) {Neural Codec Augmentation};

    \draw (xorg) [out=0, in=180] to (codec);
    \draw (xwm) [out=0, in=180, transform canvas={yshift=0.5em}] to (codec);

    \node[model, above=5em of codec] (wm) {WM Detector};

    \draw[transform canvas={xshift=-1em}] (codec) -> (wm);
    \draw[transform canvas={xshift=1em}] (codec) -> (wm);

    \node[loss, above=1em of wm] (wmloss) {WM Loss: $\mathcal{L}_\text{W}$};

    \draw (wm) -> (wmloss);
    
    \node [draw, dashed, rounded corners, inner sep=6, cyan, fit = (discriminator) (floss) (gloss) (melloss) ] (recon_ob) {};

    \node [draw, dashed, rounded corners, inner sep=6, cyan, fit = (wm) (wmloss) ] (wm_ob) {};   

    \draw[-] (xorg) to (encoder);
    \draw[-] (encoder) to (quantizer);
    \draw[-] (quantizer) to (decoder);
    \draw(decoder) to (xwm);

    \node [draw, dashed, rounded corners, inner sep=3, cyan, fit = (encoder) (quantizer) (decoder) ] (wm_model) {}; 

    \node[label, above=0.5em of xwm] (model) {Watermark Generator };
    \node[label, left=7em of model] {Reconstruction Objective};
    \node[label, right=5em of model] (wmob) {Watermarking Objective}; 

    \node[loss, right=5em of wmloss, yshift=1em] (ex1) {Loss Function};
    \node[frozen, below=1em of ex1] (ex2) {Frozen};
    \node[model, below=1em of ex2] (ex3) {Optimizable};

    \node [draw, dashed, rounded corners, inner sep=3, cyan, fit = (ex1) (ex2) (ex3) ] (legend) {}; 

    \node[label, right=3em of wmob] (legend_label) {Legend};

\end{tikzpicture}
}
\caption{Training setup.}
\label{fig:training}
\end{figure*}

\section{Methods}

% \subsection{Threat model}

% In our threat model service providers utilize zero-bit post-hoc watermarking models to distinguish synthetic speech recordings from human speech recordings with the corresponding watermark detectors. The general public does not have access to the watermarking model nor the detector, but they have access to the binary identification results from the detector.

% Watermarked audio is transmitted on the internet with DSP and neural audio codecs. Furthermore, malicious actors try to discard watermarks with neural codec resynthesis. These operations may discard the watermarks resulting in incorrect classification results. To mitigate vulnerabilities service providers use training-time channel augmentation with neural codecs. 

% Players
% Aims
% Capabilities
% Access

% Meta text

We train identical watermarking models with channel augmentation through specific neural codecs. Then we evaluate their performance against unseen attacks, both other neural codecs and DSP attacks.

\subsection{Training}

The training setup is illustrated in Figure~\ref{fig:training}. We fine-tune a pretrained neural audio codec (DAC) as a zero-bit watermark generator together with a watermark detector (AASIST)~\cite{jung2022aasist}, and the standard (DAC) adversarial discriminator. During training, original audio $x_\text{org}$ is fed through the frozen encoder and quantizer of the watermark generator, $G$. The unfrozen decoder reconstructs the audio resulting in $x_\text{wm} = G(x_\text{org})$. Encoder and quantizer are frozen, with the aim of maintaining codebook backward compatibility. This is useful in case the codes are already in use for downstream tasks, such as audio codec language models.

The watermark generator is trained with a multi-objective loss. In addition to the standard DAC training objectives which are multi-scale mel-spectrogram loss $\mathcal{L}_\text{Mel}$, least squares GAN loss $\mathcal{L}_\text{Adv}$ and a discriminator feature matching loss $\mathcal{L}_\text{Fm}$~\cite{dac}, we utilize a watermarking objective $\mathcal{L}_\text{W}$. This objective is also a least-squares GAN loss function. We use a target score of one for unwatermarked samples and zero for watermarked samples in accordance with original AASIST setup~\cite{jung2022aasist}. Both generator and watermark detector, $D$, are trained to optimize $\mathcal{L}_\text{W}$ under the collaborative training framework i.e., the watermark generator is tasked to help the detector~\cite{juvela2024-collaborative-watermarking, juvela2025audio}. Before reaching the detector, the original and watermarked audio pass through channel augmentation, $C(\cdot)$, with a separate frozen neural audio codec. The watermark detection loss is defined as
\begin{equation}
    \mathcal{L}_\text{W} = \mathbb{E}[(1-D(C(x_\text{org})))^2 + D(C(x_\text{wm}))^2]\label{eq:wmgan}.
\end{equation}
The total loss for the watermark generator then becomes 
\begin{equation}
\mathcal{L}_\text{G} = \lambda_\text{Mel}\mathcal{L}_\text{Mel} + \lambda_\text{Adv}\mathcal{L}_\text{Adv} + \lambda_\text{Fm}\mathcal{L}_\text{Fm} + \lambda_\text{W}\mathcal{L}_\text{W}.
\end{equation}
We used values $\lambda_\text{Mel} = 15$, $\lambda_\text{Adv} = 1$, $\lambda_\text{Fm} = 2$ and $\lambda_\text{W} = 1$ for the loss weight coefficients, inherited from the official DAC implementation. The standard DAC discriminator is still fine-tuned with a least squares objective $\mathcal{L}_\text{Disc}$.

%Make more concise
Channel augmentation with neural audio codecs includes a non-differentiable quantization step. These codecs typically rely on the straight-through estimator (STE)~\cite{bengio2013estimating} for their own training. However, the official releases of many neural audio codecs implement the STE operation only in training mode, which can simultaneously enable codebook updates or quantizer dropouts. For this study we modify the official source code for EnCodec, SpeechTokenizer and Mimi to enable end-to-end backpropagation with frozen models, which corresponds to real-world attacks. SNAC and DAC enable this out of the box. Our neural codec augmentation code is open-sourced\footnote{https://github.com/ljuvela/DAREA} and available for future watermarking research.
%These implementations are open-sourced and can be used for future research. %

% During the forward pass STE treats the quantization of the input variable $x$ like an offset by a constant vector
% \begin{equation}
%     \hat{x} = x + \mathrm{sg}[ q(x) -x],
% \end{equation}
% where $\hat{x}$ stands for the output variable, $q(\cdot)$ for the quantization function and $\mathrm{sg}[\cdot]$ for the stop gradient operation.

% During the backward pass this copies the gradients of the quantized variable $\hat{x}$ to the pre-quantization variable $x$, because the gradient of the constant vector is zero:

% \begin{equation}
% \frac{\partial \mathcal{L}}{\partial x} = \frac{\partial \mathcal{L}}{\partial \hat{x}} 
% \frac{\partial \hat{x}}{\partial x} = \frac{\partial \mathcal{L}}{\partial \hat{x}}\left( \frac{\partial x}{\partial x} + \frac{\partial \mathrm{sg}[ q(x) -x]}{\partial x} \right) = \frac{\partial \mathcal{L}}{\partial \hat{x}} .
% \end{equation}

% \subsection{Evaluation}

% During evaluation watermarked and original audio are passed through various neural audio codecs separately to evaluate the watermarking model's robustness against unseen neural audio codecs. Evaluation is also extended to classical DSP attacks. Evaluation also includes mean opinion score (MOS) testing to quantify the quality of watermarked audio samples in comparison to original samples. 

\section{Experiments}

\subsection{Watermarking Model and Datasets}

As a novelty, we use the official implementation of DAC~\cite{dac} as the watermark generator with a sample rate of 24\,kHz, stride of 320 samples, 32 codebooks with 1024 entries each, resulting in a bitrate of 24\,kbps. Because the official pre-trained checkpoint does not contain the weights of the GAN discriminators, we trained a base DAC model on LJSpeech~\cite{ljspeech17}, VCTK~\cite{Yamagishi2019-VCTK}, and LibriTTS-R~\cite{koizumi23_interspeech} (clean-train-100) datasets. The model was trained for one million iterations using the default training recipe and a batch size of 4 to fit on a single A100 GPU. The obtained model weights were used to initialize the training of watermarking models in our experiments.

For the watermark detector, we use the pre-trained AASIST model~\cite{jung2022aasist}. The key difference from the original implementation is that we disable batch normalization and dropout layers. Removing batch normalization has been found to boost adversarial training in previous research~\cite{juvela2024-collaborative-watermarking, wang2022removing}.

The dataset we use in our experiments is LibriTTS-R, which is a neurally restored version of the original LibriTTS corpus~\cite{koizumi23_interspeech}. The watermarking model is fine-tuned on the train-clean-100 partition of the dataset. Validation and testing were performed on the dev-clean and test-clean partitions. The native sampling rate of LibriTTS-R is 24\,kHz, which matches our watermarking architecture. 

\subsection{Channel Augmentation variants}

We train 7 identical watermarking models with different channel augmentation variants. One without any augmentation, 5 with neural codec augmentation with one specific codec and bitrate at a time, and one model variant robust against two neural codecs simultaneously. The single codec variants include: EnCodec at 6\,kbps, DAC at 8\,kbps, SpeechTokenizer at 4\,kbps, Mimi at 1.1\,kbps, and SNAC at 0.98\,kbps. The dual variant consisted of Mimi at 1.1\,kbps and SNAC at 0.98\,kbps. It was chosen after the initial experiments, as the codecs showed promising and complementary generalization. 

In the single codec setup all of the mini-batches pass through the same channel augmentation, but in the dual setup the channel augmentation variant is randomly chosen between the two options for each mini-batch. For EnCodec, DAC, SNAC and Mimi, their native sampling rate 24\,kHz matched LibriTTS-R. For channel augmentation with SpeechTokenizer, audio was resampled to 16\,kHz using Torchaudio and resampled back to 24\,kHz afterwards.

EnCodec and DAC are chosen, because they are widely used and easily available for the general public. SpeechTokenizer shares the EnCodec architecture, but has been trained with semantic distillation. Similarly, Mimi utilizes semantic distillation and it reaches a very low bitrate while maintaining high perceptual quality. Finally, SNAC was chosen, because it also operates at a very low bitrate with high quality through its hierarchical temporal tokenization. SNAC also uniquely includes a stochastic element in its decoder, while the other codecs are deterministic.

\subsection{Training details}

The watermarking model is fine-tuned for 40,000 iterations on a batch size of 8. For mini-batch training all samples are cropped or padded to 1 second. Validation is performed with a batch size of 4 and validation samples were padded or cropped to 5 seconds. 

All of the training used the AdamW optimizer with $\beta_1 = 0.8, \beta_2 = 0.99$. Optimizer and learning rate scheduler states were reset after initializing each model using the pre-trained model weights. We used NVIDIA A100 GPUs for training, and each model was trained on a single GPU. 

The best model version was determined on a 250 sample subset of the dev-clean partition based on the best equal-error-rate (EER). During validation the watermarked audio and the reference audio are passed through the same channel augmentation as used for training.

\begin{table*}[ht]
\setlength{\tabcolsep}{2pt}
\caption{Experiments TPR@1\%FPR (Higher is Better).}\label{tab:tpr}
\begin{threeparttable}
\centering
\resizebox{.9\textwidth}{!}{
\begin{tabular}{cccccccccc}
\toprule 
&  \multicolumn{7}{c}{Augmentations} & \multicolumn{2}{c}{Baselines}

\\
\hline  \rule{0pt}{3ex}                          Attacks $\downarrow$ &    No Codec     &     EnCodec     &       DAC       & SpeechTokenizer &      Mimi       &      SNAC       &  Mimi and SNAC  &    AudioSeal    &     WavMark    \\ 
\midrule
         No Codec          & \cellcolor[rgb]{0.41, 0.67, 0.81} 100.00 & \cellcolor[rgb]{0.98, 0.76, 0.65} 21.00 & \cellcolor[rgb]{0.58, 0.78, 0.87} 89.20 & \cellcolor[rgb]{0.48, 0.71, 0.84} 95.80 & \cellcolor[rgb]{0.54, 0.75, 0.86} 92.10 & \cellcolor[rgb]{0.58, 0.78, 0.87} 89.30 & \cellcolor[rgb]{0.47, 0.71, 0.84} 96.80 & \cellcolor[rgb]{0.41, 0.67, 0.81} 100.00 & \cellcolor[rgb]{0.41, 0.67, 0.81} 100.00\\ 
         \hline
      EnCodec 6kbps        & \cellcolor[rgb]{0.90, 0.53, 0.42} 1.40 & \cellcolor[rgb]{0.42, 0.68, 0.82} 99.70 & \cellcolor[rgb]{0.93, 0.59, 0.46} 5.70 & \cellcolor[rgb]{0.98, 0.78, 0.67} 22.40 & \cellcolor[rgb]{0.44, 0.69, 0.83} 98.10 & \cellcolor[rgb]{0.59, 0.78, 0.87} 88.90 & \cellcolor[rgb]{0.44, 0.69, 0.83} 97.90 & \cellcolor[rgb]{0.87, 0.92, 0.95} 62.70 & \cellcolor[rgb]{0.90, 0.51, 0.41} 0.52\\ 
      EnCodec 12kbps       & \cellcolor[rgb]{0.90, 0.51, 0.41} 0.50 & \cellcolor[rgb]{0.42, 0.68, 0.82} 99.70 & \cellcolor[rgb]{0.98, 0.80, 0.71} 24.80 & \cellcolor[rgb]{0.94, 0.95, 0.96} 54.20 & \cellcolor[rgb]{0.43, 0.68, 0.82} 98.70 & \cellcolor[rgb]{0.52, 0.74, 0.85} 93.10 & \cellcolor[rgb]{0.44, 0.69, 0.83} 97.90 & \cellcolor[rgb]{0.43, 0.68, 0.82} 99.20 & \cellcolor[rgb]{0.91, 0.55, 0.43} 2.51\\ 
        DAC 8kbps          & \cellcolor[rgb]{0.90, 0.53, 0.42} 1.40 & \cellcolor[rgb]{0.94, 0.60, 0.47} 6.80 & \cellcolor[rgb]{0.42, 0.68, 0.82} 99.80 & \cellcolor[rgb]{0.87, 0.92, 0.95} 63.60 & \cellcolor[rgb]{0.48, 0.71, 0.84} 95.60 & \cellcolor[rgb]{0.62, 0.79, 0.88} 86.70 & \cellcolor[rgb]{0.44, 0.69, 0.83} 98.00 & \cellcolor[rgb]{0.41, 0.67, 0.81} 100.00 & \cellcolor[rgb]{0.42, 0.68, 0.82} 99.37\\ 
  SpeechTokenizer 4kbps    & \cellcolor[rgb]{0.90, 0.53, 0.42} 1.20 & \cellcolor[rgb]{0.91, 0.54, 0.42} 2.20 & \cellcolor[rgb]{0.91, 0.54, 0.42} 2.00 & \cellcolor[rgb]{0.43, 0.68, 0.82} 98.90 & \cellcolor[rgb]{0.99, 0.83, 0.75} 28.10 & \cellcolor[rgb]{0.92, 0.95, 0.96} 57.00 & \cellcolor[rgb]{0.87, 0.92, 0.95} 62.60 & \cellcolor[rgb]{0.91, 0.54, 0.42} 2.20 & \cellcolor[rgb]{0.91, 0.54, 0.42} 2.20\\ 
       Mimi 1.1kbps        & \cellcolor[rgb]{0.90, 0.53, 0.42} 0.90 & \cellcolor[rgb]{0.90, 0.53, 0.42} 1.10 & \cellcolor[rgb]{0.90, 0.53, 0.42} 0.90 & \cellcolor[rgb]{0.91, 0.54, 0.42} 1.70 & \cellcolor[rgb]{0.47, 0.71, 0.84} 96.70 & \cellcolor[rgb]{0.97, 0.94, 0.92} 45.00 & \cellcolor[rgb]{0.65, 0.81, 0.89} 83.60 & \cellcolor[rgb]{0.90, 0.53, 0.42} 1.30 & \cellcolor[rgb]{0.90, 0.51, 0.41} 0.21\\ 
      SNAC 0.98kbps        & \cellcolor[rgb]{0.90, 0.51, 0.41} 0.50 & \cellcolor[rgb]{0.90, 0.53, 0.42} 0.90 & \cellcolor[rgb]{0.90, 0.53, 0.42} 1.10 & \cellcolor[rgb]{0.91, 0.54, 0.42} 1.70 & \cellcolor[rgb]{0.91, 0.54, 0.42} 2.30 & \cellcolor[rgb]{0.47, 0.71, 0.84} 96.80 & \cellcolor[rgb]{0.69, 0.83, 0.91} 80.00 & \cellcolor[rgb]{0.90, 0.53, 0.42} 1.00 & \cellcolor[rgb]{0.90, 0.53, 0.42} 1.05\\ 
 NanoCodec 1.89kbps   & \cellcolor[rgb]{0.91, 0.54, 0.42} 1.80 & \cellcolor[rgb]{0.91, 0.54, 0.42} 1.90 & \cellcolor[rgb]{0.91, 0.54, 0.42} 1.90 & \cellcolor[rgb]{0.91, 0.55, 0.43} 2.40 & \cellcolor[rgb]{0.77, 0.87, 0.93} 73.70 & \cellcolor[rgb]{0.81, 0.89, 0.94} 70.90 & \cellcolor[rgb]{0.73, 0.85, 0.92} 77.20 & \cellcolor[rgb]{0.90, 0.53, 0.42} 0.90 & \cellcolor[rgb]{0.90, 0.51, 0.41} 0.42\\ 
    \hline
        MP3 32kbps         & \cellcolor[rgb]{0.98, 0.91, 0.88} 40.60 & \cellcolor[rgb]{0.95, 0.63, 0.50} 8.60 & \cellcolor[rgb]{0.99, 0.88, 0.81} 33.50 & \cellcolor[rgb]{0.97, 0.96, 0.96} 49.20 & \cellcolor[rgb]{0.91, 0.94, 0.96} 57.10 & \cellcolor[rgb]{0.89, 0.93, 0.95} 60.20 & \cellcolor[rgb]{0.61, 0.79, 0.88} 87.20 & \cellcolor[rgb]{0.41, 0.67, 0.81} 100.00 & \cellcolor[rgb]{0.41, 0.67, 0.81} 100.00\\ 
      OGG Vorbis Q2        & \cellcolor[rgb]{0.95, 0.63, 0.50} 8.80 & \cellcolor[rgb]{0.93, 0.58, 0.46} 5.30 & \cellcolor[rgb]{0.95, 0.62, 0.49} 8.10 & \cellcolor[rgb]{0.98, 0.77, 0.66} 21.50 & \cellcolor[rgb]{0.88, 0.93, 0.95} 61.70 & \cellcolor[rgb]{0.98, 0.93, 0.90} 42.70 & \cellcolor[rgb]{0.69, 0.83, 0.91} 80.00 & \cellcolor[rgb]{0.41, 0.67, 0.81} 100.00 & \cellcolor[rgb]{0.65, 0.81, 0.89} 83.04\\ 
     OGG Opus 32kbps       & \cellcolor[rgb]{0.90, 0.94, 0.96} 59.30 & \cellcolor[rgb]{0.95, 0.63, 0.50} 8.60 & \cellcolor[rgb]{0.97, 0.97, 0.97} 50.20 & \cellcolor[rgb]{0.92, 0.95, 0.96} 55.60 & \cellcolor[rgb]{0.83, 0.90, 0.94} 69.00 & \cellcolor[rgb]{0.77, 0.87, 0.93} 73.50 & \cellcolor[rgb]{0.53, 0.75, 0.86} 92.40 & \cellcolor[rgb]{0.41, 0.67, 0.81} 100.00 & \cellcolor[rgb]{0.41, 0.67, 0.81} 100.00\\ 
        MP3 64kbps         & \cellcolor[rgb]{0.43, 0.68, 0.82} 98.90 & \cellcolor[rgb]{0.95, 0.64, 0.51} 9.40 & \cellcolor[rgb]{0.65, 0.81, 0.89} 83.00 & \cellcolor[rgb]{0.63, 0.80, 0.89} 85.90 & \cellcolor[rgb]{0.62, 0.79, 0.88} 86.40 & \cellcolor[rgb]{0.61, 0.79, 0.88} 86.90 & \cellcolor[rgb]{0.46, 0.70, 0.83} 96.90 & \cellcolor[rgb]{0.41, 0.67, 0.81} 100.00 & \cellcolor[rgb]{0.41, 0.67, 0.81} 100.00\\ 
      OGG Vorbis Q3        & \cellcolor[rgb]{0.99, 0.88, 0.81} 33.00 & \cellcolor[rgb]{0.93, 0.58, 0.46} 5.30 & \cellcolor[rgb]{0.97, 0.71, 0.59} 15.80 & \cellcolor[rgb]{0.87, 0.92, 0.95} 63.40 & \cellcolor[rgb]{0.59, 0.78, 0.87} 88.40 & \cellcolor[rgb]{0.91, 0.94, 0.96} 57.20 & \cellcolor[rgb]{0.60, 0.78, 0.88} 88.00 & \cellcolor[rgb]{0.41, 0.67, 0.81} 100.00 & \cellcolor[rgb]{0.44, 0.69, 0.83} 98.32\\ 
     OGG Opus 64kbps       & \cellcolor[rgb]{0.43, 0.68, 0.82} 99.00 & \cellcolor[rgb]{0.96, 0.65, 0.52} 10.50 & \cellcolor[rgb]{0.67, 0.82, 0.90} 81.30 & \cellcolor[rgb]{0.67, 0.82, 0.90} 81.80 & \cellcolor[rgb]{0.64, 0.80, 0.89} 85.10 & \cellcolor[rgb]{0.71, 0.84, 0.91} 78.20 & \cellcolor[rgb]{0.48, 0.71, 0.84} 95.50 & \cellcolor[rgb]{0.41, 0.67, 0.81} 100.00 & \cellcolor[rgb]{0.41, 0.67, 0.81} 100.00\\ 
         \hline
      MIT RIR Reverb       & \cellcolor[rgb]{0.96, 0.68, 0.55} 12.70 & \cellcolor[rgb]{0.91, 0.55, 0.43} 3.00 & \cellcolor[rgb]{0.98, 0.78, 0.68} 23.30 & \cellcolor[rgb]{0.94, 0.61, 0.48} 7.70 & \cellcolor[rgb]{0.95, 0.62, 0.49} 8.10 & \cellcolor[rgb]{0.93, 0.58, 0.46} 4.70 & \cellcolor[rgb]{0.98, 0.81, 0.72} 25.40 & \cellcolor[rgb]{0.67, 0.82, 0.90} 81.70 & \cellcolor[rgb]{0.41, 0.67, 0.81} 100.00\\ 
       Musan Noise         & \cellcolor[rgb]{0.63, 0.80, 0.89} 85.90 & \cellcolor[rgb]{0.98, 0.76, 0.65} 20.70 & \cellcolor[rgb]{0.85, 0.91, 0.95} 65.40 & \cellcolor[rgb]{0.97, 0.97, 0.96} 49.40 & \cellcolor[rgb]{0.51, 0.73, 0.85} 94.10 & \cellcolor[rgb]{0.76, 0.87, 0.92} 74.60 & \cellcolor[rgb]{0.59, 0.78, 0.87} 88.50 & \cellcolor[rgb]{0.42, 0.68, 0.82} 99.90 & \cellcolor[rgb]{0.44, 0.69, 0.83} 98.32\\ 
       Lowpass 4kHz        & \cellcolor[rgb]{0.43, 0.68, 0.82} 98.80 & \cellcolor[rgb]{0.92, 0.56, 0.44} 3.60 & \cellcolor[rgb]{0.92, 0.95, 0.96} 56.50 & \cellcolor[rgb]{0.54, 0.75, 0.86} 91.60 & \cellcolor[rgb]{0.98, 0.91, 0.88} 40.60 & \cellcolor[rgb]{0.99, 0.88, 0.82} 34.10 & \cellcolor[rgb]{0.83, 0.90, 0.94} 69.40 & \cellcolor[rgb]{0.41, 0.67, 0.81} 100.00 & \cellcolor[rgb]{0.41, 0.67, 0.81} 100.00\\ 
      Highpass 500Hz       & \cellcolor[rgb]{0.41, 0.67, 0.81} 100.00 & \cellcolor[rgb]{0.99, 0.83, 0.75} 27.40 & \cellcolor[rgb]{0.59, 0.78, 0.87} 88.60 & \cellcolor[rgb]{0.44, 0.69, 0.83} 97.70 & \cellcolor[rgb]{0.54, 0.75, 0.86} 91.80 & \cellcolor[rgb]{0.61, 0.79, 0.88} 87.30 & \cellcolor[rgb]{0.44, 0.69, 0.83} 97.90 & \cellcolor[rgb]{0.41, 0.67, 0.81} 100.00 & \cellcolor[rgb]{0.41, 0.67, 0.81} 100.00\\ 
          \hline
 Two Random Neural Codecs  & \cellcolor[rgb]{0.90, 0.53, 0.42} 1.00 & \cellcolor[rgb]{0.96, 0.69, 0.57} 14.40 & \cellcolor[rgb]{0.94, 0.60, 0.47} 6.50 & \cellcolor[rgb]{0.94, 0.60, 0.47} 6.90 & \cellcolor[rgb]{0.99, 0.87, 0.80} 32.70 & \cellcolor[rgb]{0.98, 0.94, 0.91} 44.30 & \cellcolor[rgb]{0.96, 0.96, 0.97} 50.90 & \cellcolor[rgb]{0.94, 0.61, 0.48} 7.20 & \cellcolor[rgb]{0.90, 0.53, 0.42} 1.05\\ 
Three Random Neural Codecs & \cellcolor[rgb]{0.90, 0.53, 0.42} 1.50 & \cellcolor[rgb]{0.93, 0.58, 0.46} 5.00 & \cellcolor[rgb]{0.90, 0.53, 0.42} 1.50 & \cellcolor[rgb]{0.91, 0.54, 0.42} 2.20 & \cellcolor[rgb]{0.97, 0.71, 0.59} 15.90 & \cellcolor[rgb]{0.98, 0.80, 0.71} 24.40 & \cellcolor[rgb]{0.98, 0.78, 0.67} 21.90 & \cellcolor[rgb]{0.91, 0.54, 0.42} 1.80 & \cellcolor[rgb]{0.90, 0.51, 0.41} 0.11\\ 
\bottomrule
\end{tabular}
}

\end{threeparttable}
\end{table*}

\subsection{Evaluation}

Evaluation is done on a 1,000 sample subset of the test-clean partition, with the maximum file size being cropped to 10 seconds. During evaluation reference audio and watermarked audio pass through a specific attack before reaching the detector. Detection robustness is then calculated for each channel augmentation variant.

Evaluation is performed on all of the aforementioned neural audio codecs. Furthermore, we extended neural codec attacks to EnCodec at 12\,kbps and NanoCodec at 1.89\,kbps. Note that for NanoCodec, we include appropriate resampling to and from 22.050\,kHz. We also evaluated robustness against traditional codecs, MP3, Ogg Vorbis and Opus, at various bitrates as well as against additive noise~\cite{musan2015}, reverberation~\cite{rir}, lowpass and highpass filtering. We also include sequential multiple resynthesis with randomly sampled permutation of neural codecs from EnCodec (at both 6\,kbps and 12\,kbps), DAC, SpeechTokenizer, Mimi and SNAC.

AudioSeal and WavMark are state-of-the art post-hoc watermarking models and we use them for baseline robustness comparison~\cite{audioseal, chen2023wavmark}. We downsample recordings to 16\,kHz using Torchaudio to match the native sampling rate of the models. For listening tests we keep the original lost frequency band and combine it with the watermarked signal, which is upsampled back to 24\,kHz.  Important difference is that AudioSeal attempts to watermark each sampling point of the audio signal with localized watermarking and WavMark operates on 1 second long segments, whereas our models operate on the whole recording with global watermarks. 

For the WavMark comparison, we watermark each consecutive 1-second segment with a known 32-bit code. During detection we average the decoded bit probabilities across consecutive 1-second segments (we assume known locations), threshold at 0.5 to obtain the final 32-bit message, and use the fraction of matching bits as the global watermark probability score. AudioSeal, on the other hand, outputs a per-sample probability. We follow the official implementation and take the fraction of sample scores above 0.5 as the score.

\section{Results}

\subsection{Detection results}

Results of the experiments are reported with the true positive rate at 1 percent false positive rate (TPR@1FPR) in Table~\ref{tab:tpr}. Although AASIST uses natural samples as the positive class, the TPR@1FPR calculations use watermarked samples as the positive class.

Results show that there are large differences in watermarking robustness depending on the neural codec used for channel augmentation. EnCodec and DAC show poor generalization to other neural codecs, however SpeechTokenizer already performs better. Mimi and SNAC showcase significantly better generalization to various attacks when used individually, however combining the two results in the broadest generalization to neural codec and DSP attacks. 

Dual Mimi and SNAC augmentation performs very well against EnCodec, DAC and classical DSP codecs. However, performance decreases for SpeechTokenizer and NanoCodec. We can also observe moderate robustness against additive noise and filtering. Reverberation and multiple resynthesis remain challenging for our watermarking models. Interestingly, the dual model does not achieve above 90\% robustness against neither SNAC nor Mimi. Furthermore, Mimi and SNAC augmentation generalize poorly to each other. This suggests that channel augmentation with the two codecs could encourage contradictory gradients to the signal.

AudioSeal has been trained to be robust against a multitude of attacks, including EnCodec and MP3 resynthesis, filtering, additive noise and resampling among others~\cite{audioseal}. However, AudioSeal catastrophically fails against attacks with SpeechTokenizer, Mimi, SNAC and NanoCodec, although it does survive resynthesis with DAC. WavMark has been similarly trained to survive particular DSP attacks~\cite{chen2023wavmark} but it fails catastrophically with all neural codec attacks except for DAC resynthesis. AudioSeal and WavMark results are broadly in line with previous robustness experiments \cite{shallow}. Our models achieve broader robustness with a much more restricted number of augmentations. Future experiments should investigate how augmentation with Mimi and SNAC works with localized watermarking models such as AudioSeal.

\subsection{Quality evaluation}

We evaluated audio quality with a comparative mean opinion score (CMOS) test, using original LibriTTS-R samples as references against watermarked ones. We also compared the natural samples against AudioSeal, WavMark, and our pretrained DAC model.

Subjective evaluation set consisted of 29 utterances between 3 and 9.5 seconds from our robustness test set. These utterances were rendered with all of the system variants and normalized to equal loudness (-24 LUFS) together with the original samples. For automated evaluation, we also calculated PESQ scores~\cite{pesq} on the whole robustness test set.

The listening test was conducted via the crowd-sourcing platform Prolific. Each participant rated 5 randomly selected utterances from the subjective evaluation set for each of the system configurations, resulting in 50 samples rated for each participant. Utterances were drawn independently for each system to reduce listener fatigue.

Participants were asked to rate the quality of the samples on a Likert scale from -3 (sample A is much better) to 3 (sample B is much better) using headphones. The order of reference and target samples was randomized for each sample. The test also included attention checks to verify the quality of submissions. Altogether, 26 out of 28 participants passed the attention checks. 

For each user, we calculated the mean rating they gave for each system. Then we analyzed these aggregated ratings using Student's t-statistic. Our results present the mean rating and 95\% confidence intervals adjusted for multiple comparisons using Dunn-Bonferroni correction. User-based aggregation and a shared utterance pool were used to reduce unwanted bias and enhance the statistical robustness of MOS analysis~\cite{rosenberg17_interspeech}. 

\begin{table}[htbp]
\caption{Quality Testing. CMOS against the natural samples.}\label{tab:mos}
\centering
\begin{tabular}[t]{lcrr}
\toprule
Generator & Augmentation & CMOS $\bar{x} \pm \text{CI}$ & PESQ\\
\midrule
DAC & No Codec & $0.231 \pm 0.269$ & 4.429\\
DAC & SNAC & $0.154 \pm 0.338$ & 4.398\\
DAC & Pretrained & $0.100 \pm 0.245$ & 4.423\\
DAC & EnCodec & $0.077 \pm 0.227$ & 4.409\\
DAC & DAC & $0.000 \pm 0.354$ & 4.420\\
DAC & SpeechTokenizer & $-0.008 \pm 0.243$ & 4.408\\
DAC & Mimi & $-0.023 \pm 0.319$ & 4.395\\
DAC & Mimi + SNAC & $-0.092 \pm 0.300$ & 4.385\\
\hline
AudioSeal & See \cite{audioseal} & $-0.138 \pm 0.464$ & 4.339\\
WavMark & See \cite{chen2023wavmark} & $-0.292 \pm 0.432$ & 4.133\\
\bottomrule
\end{tabular}
\end{table}

The results are listed in Table~\ref{tab:mos} and show that all of the systems remain close to original samples in quality. This is further demonstrated by the consistently high PESQ scores. The results suggest that rendered samples are very difficult to distinguish from natural samples. We can also observe that for some of the systems users rate watermarked samples slightly higher than original LibriTTS-R samples, however this can be explained by random variation together with the GAN-objective. The results show that our DAC-based watermarking models achieve good audio quality that is comparable to natural samples and SOTA watermarking models. Note that we do not claim that our method is better than AudioSeal or WavMark, because the models have different assumptions and native sample rates, but the models have comparable quality.

\section{Discussion and Conclusions}

The results demonstrate that neural audio codecs should not be treated as a generic augmentation strategy for watermarking. Robustness against unseen neural codecs and DSP attacks is significantly influenced by the choice of the channel augmentation codec. Augmentation with Mimi and SNAC shows promising generalization across unseen neural codecs and classical distortions. This could be explained by their very low bitrate coupled with high reconstruction quality, which forces the watermarking model to embed the watermark into a more integral part of the signal.

However, the experiments also show that Mimi and SNAC augmentation fail to generalize to each other and that they perform worse against NanoCodec and SpeechTokenizer than EnCodec and DAC. This should be seen as a limitation of the current work, as it is possible that other neural codecs with similarly efficient bottlenecks remain viable attack methods against Mimi and SNAC augmentation. The results still indicate that if watermarking models are trained to survive neural resynthesis with state-of-the-art models, this likely generalizes to older lower-capacity models.

Furthermore, our models were not evaluated against neural denoisers nor vocoders, which have been identified as powerful attack vectors~\cite{shallow}. Future research is needed to create audio watermarks which remain invariant against a wider range of neural attacks without excessive quality degradation. 

\clearpage
\clearpage

\section{Acknowledgements}
We acknowledge the computational resources provided by the Aalto Science-IT project. We used Aalto AI assistant, GitHub Copilot and Google Gemini for reviewing and modifying code. All changes were implemented and validated by the authors.

\bibliographystyle{IEEEbib}
\bibliography{myRefs}

\end{document}